# Wide Field Imaging Interferometry Testbed III – metrology subsystem

Douglas B. Leviton[*a], Bradley J. Frey[a], David T. Leisawitz[a], Anthony J. Martino[a], William L. Maynard[a], Lee G. Mundy[b], Stephen A. Rinehart[c], Stacy H. Teng[b], Xiaolei Zhang[d],
[a]NASA Goddard Space Flight Center, [b]Department of Astronomy, University of Maryland,
[c]National Research Council, [d]SSAI

## ABSTRACT

In order for data products from WIIT to be as robust as possible, the alignment and mechanical positions of source, receiver, and detector components must be controlled and measured with extreme precision and accuracy, and the ambient environment must be monitored to allow environmental effects to be correlated with even small perturbations to fringe data. Relevant detailed anatomy of many testbed components and assemblies are described. The system of displacement measuring interferometers (DMI), optical encoders, optical alignment tools, optical power monitors, and temperature sensors implemented for control and monitoring of the testbed is presented.

Keywords: metrology, interferometer, absolute encoder, delay line, servo, testbed, alignment, wide-field imaging

## 1. BACKGROUND

The Wide-field Imaging Interferometry Testbed (WIIT) is a laboratory demonstrator for observational and computational techniques which allow synthetic aperture imaging of moderately wide fields of view (in astronomical terms) at angular resolutions which are higher than that possible by the diffraction limit of its individual collecting apertures[1,2]. It is a visible light analogue of the planned astronomy mission Sub-millimeter Probe of the Evolution of Cosmic Structure (SPECS).[3,4] A goal of SPECS is to be able to record images of the sky in the sub-millimeter wavelength range which are equivalent in field of view and angular resolution to the famous Hubble Deep Field (HDF) image taken across the visible spectrum by Hubble Space Telescope (HST).

The testbed's successful operation will help validate observational techniques and data analysis tools for the SPECS mission. WIIT includes a source assembly, a 0.5 m diameter, f/4.5 parabolic collimating mirror, a receiver assembly, and a detector assembly, all atop a 1.2 m wide x 3.6 m long x 0.3 m thick vibrationally-isolated optical table (Figure 1). The arrangement of a rotating source and linearly scanned receiver baseline mirrors allows the u-v plane of the synthesized testbed aperture to be sampled in a polar coordinate system. The system to provide environmental, optical, and electro-mechanical metrology needed for proper testbed function is discussed in the paper.

## 2. WIIT ANATOMY AND METROLOGY OVERVIEW

There are several categories of metrology required in WIIT for fringe data to be optimized or indeed to even make sense: 1) electro-mechanical displacement and rotation, 2) optical alignment, 3) optical path monitoring, 4) optical power monitoring, and 5) temperature sensing of air through which beams pass and temperature sensing of hardware components adjacent to beam paths. The means to implement this combination of environmental and electro-opto-mechanical sensing and the rationale for doing so are discussed in the subsequent sections. In some cases, data which illustrate those rationale are presented.

To become familiar with the requirements for each of type of metrology, it is useful to describe the anatomy of the testbed and it constituent assemblies in some detail. Let us start with an overview of what happens to light as it makes its way through the testbed.

---

* doug.leviton@gsfc.nasa.gov, phone 1 301 286-3670, fax 1 301 286-0204, NASA Goddard Space Flight Center, Code 551, Greenbelt, MD, 20771

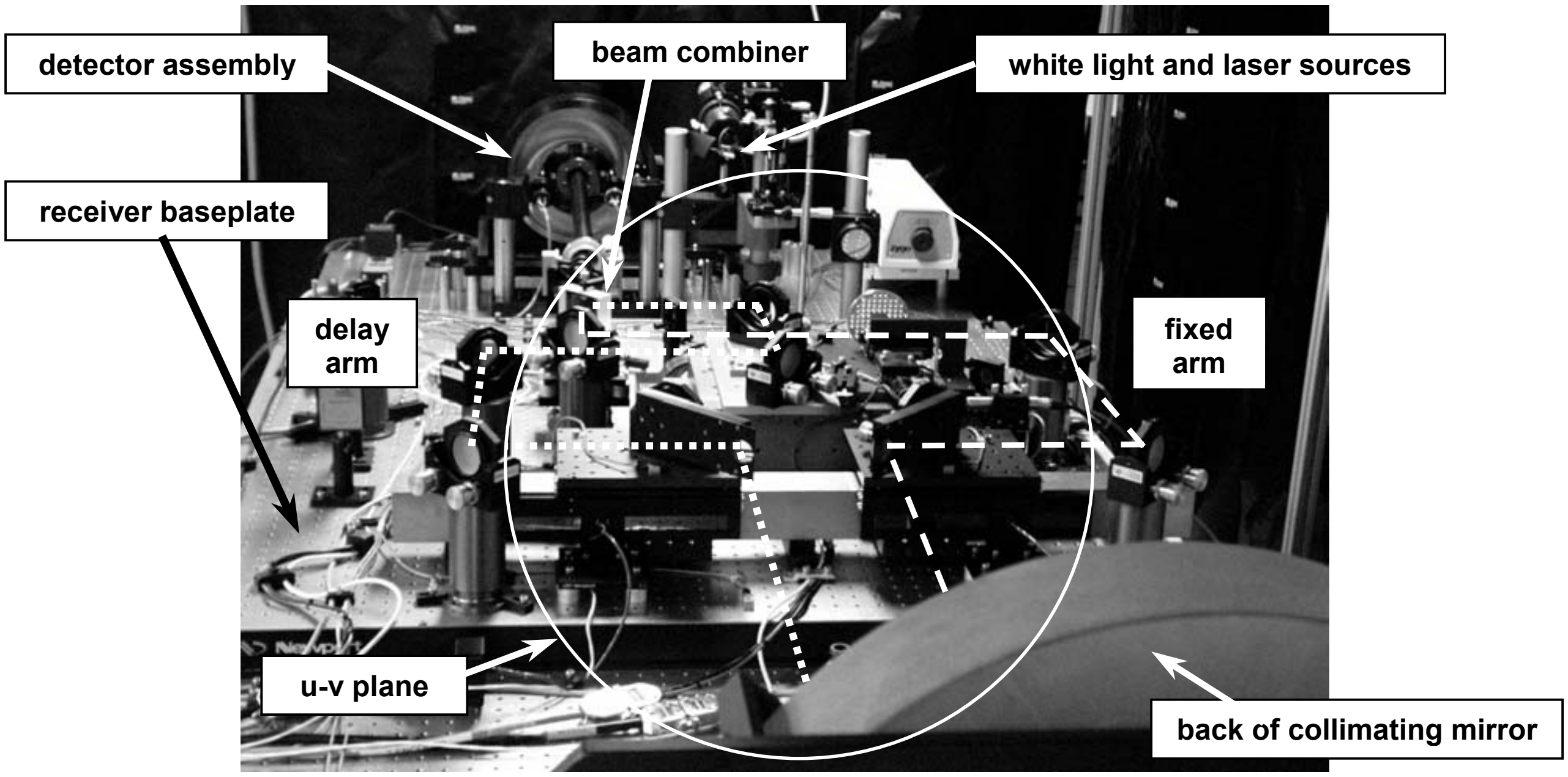

Figure 1 – overview of WIIT implementation: source assembly (laser and white light), collimating mirror, receiver (built on 1.2 m x 1.2 m optical breadboard), and detector assembly (CCD camera with long tubular baffle) all on large vibration isolation table

A source scene, illuminated with white light, filtered light, or laser light, is rotated to sweep out the angular component of the u-v plane. The collimator causes the illuminated scene to appear to the receiver to be at infinite distance. The receiver's flat baseline mirrors, which feed two arms of a Michelson interferometer, are scanned co-linearly to sample the radial component of the u-v plane. Fixed flat mirrors fold the sampled beams toward a beamsplitter/combiner. One arm contains a dihedral mirror arrangement which is scanned linearly to create optical delay (optical path difference) between the two arms, allowing spectral information of the scene to be deduced through Fourier transformation of recorded fringe data, and allowing resolution of zero path difference (ZPD) for off-axis field points (resulting in a wide field of view capability)[2,3]. The two arms are recombined at the beamsplitter where interference occurs between correlated samples of the u-v plane. The detector assembly consists of a 750 mm focal length lens and a high frame rate CCD array camera. The lens re-images the source field onto the CCD providing the imaging aspect of the testbed. The intensity of each image point is modulated as the delay line is stroked due to changes in relative phase of the two arms at the beamsplitter. This is the interferometric aspect of the testbed. The camera can be rotated in synchrony with source rotation to simulate rotation of the receiver itself with or without respect to a fixed astronomical scene.

## 3. DETAILED ANATOMY AND METROLOGY OF WIIT ASSEMBLIES

In this section, we discuss each testbed assembly in detail. Metrology requirements pertaining to each component and metrology means implemented to meet the requirements are presented. Optical alignment tools and techniques used to make the testbed work as a system are also discussed.

### 3.1 Source assembly

The source assembly is built on a high precision, motorized, indexing air bearing spindle whose rotation axis is aligned coincident with the axis of the parabolic collimating mirror (Figure 2a). The spindle carries a high precision, micrometer driven X-Y microscope stage. The stage holds a microlithographically patterned chrome-on-glass plate containing an array of over 150 interesting targets for the testbed to image (Figure 2b). The critical dimension of the smallest target features is just smaller than what the testbed should theoretically be able to resolve. The source assembly is placed in the testbed so that the target plate is in the focal plane of the parabolic collimator. The side drive micrometers on the stage axes have 50 mm of travel with 10 microns resolution. The microscope stage allows any portion of the 50 mm x 50 mm target plate to be positioned on the axis of the collimator to within about the size of that critical dimension.

The requirement on knowledge of absolute angle of source rotation has been derived previously as 0.3 arcseconds. A couple of things besides just packaging issues go into the selection of a bearing and angle sensor capable of meeting such a requirement. The mechanical runout of the bearing for the spindle axis must be exceptionally small, and the rotary encoder must be of commensurate resolution and accuracy. A numerical example will emphasize this point. If an optical encoder with a track radius of 50 mm experiences a radial shift of even 1 μm due to runout of the axis of a rotary bearing, but there is, in fact, no accompanying rotation of the axis, a read head in the encoder may errantly interpret the runout as a rotation of $2 \times 10^{-5}$ radians or 4 arcseconds, regardless of the intrinsic accuracy of the encoder. In practice, optical encoders usually employ at least two diametrically opposed read heads, because the average of the readings from the two heads algebraically cancels the effect of runout error. Still, a runout free bearing is highly desirable.

A motorized, air bearing spindle with 0.025 μm rms runout to rotates the source scene. For the encoder in the example above, this runout gives a maximum error of 0.1 arcseconds. The spindle's step resolution is 0.2 arcseconds. The encoder on the air bearing spindle is a rotary version of NASA's new technology, absolute, pattern recognition encoder.[5,6,7] The encoder has a track radius of roughly 60 mm, a fiducial spacing of 0.08789 degrees, an associated angular resolution of 0.02 arcseconds (using 10X magnification for the encoder scale), and a measured accuracy versus a calibrated true square of better than 0.1 arcseconds. Two diametrically opposed read stations are used to cancel possible bearing runout errors and to improve accuracy and resolution over that achievable with a single read station (Figure 2c).

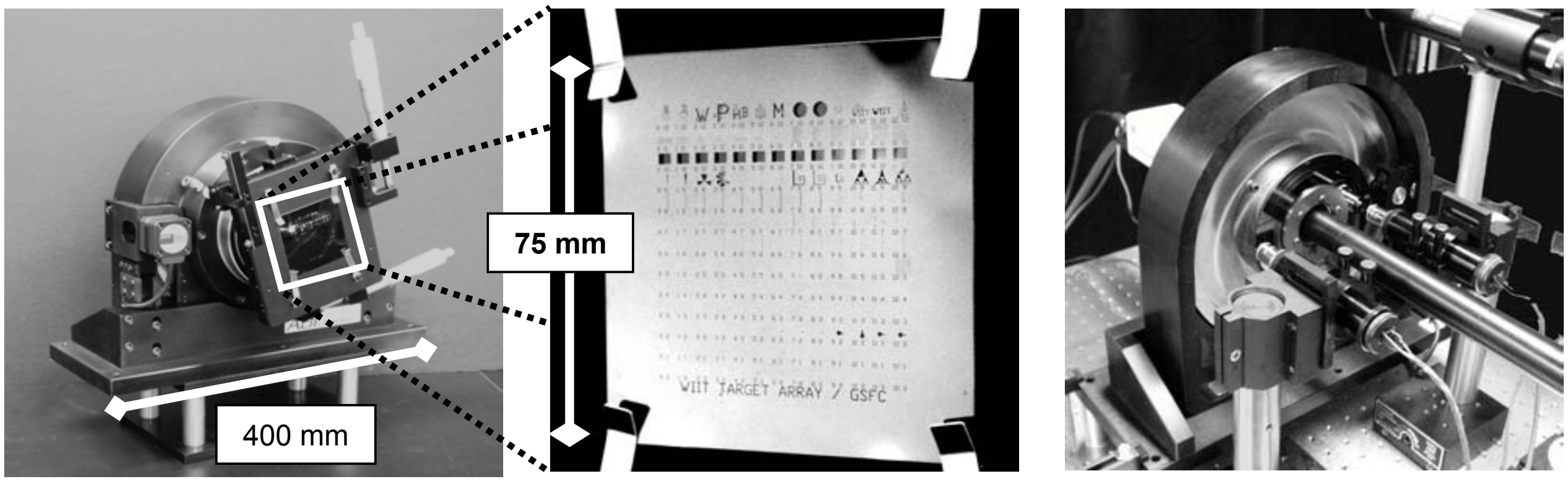

a) b) c)

Figure 2 – a) source assembly showing target array plate in X-Y microscope stage; b) detailed view of glass target array plate; c) encoder configuration on back side of spindle is the same as for detector assembly (shown)

## 3.2 Receiver assembly

### 3.2.1 Baseline

The receiver is an arrangement of two series of high accuracy flat mirrors (1/20 of a visible wave PV) coated with bare aluminum, comprising two arms of a Michelson interferometer. The flats are mounted in compact gimbals having sub-arcsecond level tip and rotation adjustments. Each arm begins with a diagonal mirror of 25 mm projected diameter to the collimated beam picking off portions of the collimated beam in accordance with the desired sampling of the u-v plane. The diagonal receiver mirrors travel along a common baseline, their mounts fastened to the sliding carriages of a custom, dual-carriage, air bearing rail with outstanding pitch, roll, and yaw performance of the traveling carriages.

The requirement for pitch, roll, and yaw of the diagonal mirrors as they travel along the baseline is derived from the goal to have the two images of the same source point, viewed by the separate baseline mirrors, land within one pixel of each other on the CCD without compensating angular adjustments to the baseline mirrors. This would give maximum white light fringe contrast for any pairing of baseline mirror positions at ZPD. Each CCD pixel corresponds to about 2.5 arcseconds of field of view. Because beams are deflected by a rotated flat mirror by twice the angle of rotation, this translates to a differential pointing requirement of the baseline mirrors of just less than 1.3 arcseconds.

The requirement on knowledge of the absolute position of each receiver mirror along the baseline had once been derived to be equal to about one-third of the typical wavelength of interest in the testbed (roughly 0.6 μm), or 0.2 μm.[2] This assumed that there would be sources in the field of view which would serve as phase references for resolving ZPD for different baselines. If no such sources are available, the situation is complicated and will be treated in Section 3.2.2.

Initially, baseline mirrors were to be positioned using commercial, DC servo translation stages with 0.1 μm resolution. However, while straightness and flatness of travel of these mechanical stages may have met specifications (we never had occasion to check), their pitch, roll, and yaw performance out of the box was found to be highly inadequate, exceeding 30 arcseconds in at least pitch and yaw. This type of performance would make it impossible to keep the two images of the same point source coincident on the CCD, effectively guaranteeing that required interference would seldom occur!

The solution to the pitch, roll, and yaw performance issue, was to engineer the custom, dual carriage air slide to carry the baseline mirrors, and use the mechanical stages (already a substantial financial investment) as actuators for the air carriages with servo control (Figure 3). Position knowledge is gained through NASA's linear, absolute, encoders (Figure 4). Encoders on the baseline carriages use no imaging optics, but instead rely on shadowing of the scale pattern onto CCD's in close proximity. In this configuration, resolution of the linear encoders is about 0.05 μm, and uncalibrated accuracy is about 0.1 μm per 100 mm at 21 C. A mechanical link between the stage and the air carriage consists of a ball tipped finger mounted to the stage, a flat, hardened steel contact surface on the air carriage, and a steel spring leaf to keep the two in contact without allowing errant motion of the mechanical stage to perturb the body orientation of the air carriage. The mechanical stages have 200 mm travel. The air carriages allow a maximum travel of just over 250 mm, as do the encoders. In order to avoid thermal drifts in calibration of every testbed encoder from instabilities which might occur with socketed CCD's on printed circuit boards in separate camera housings, CCD's are removed from their sockets and epoxied to stable, rigidly mounted assembly parts and then pigtailed out to their sockets.

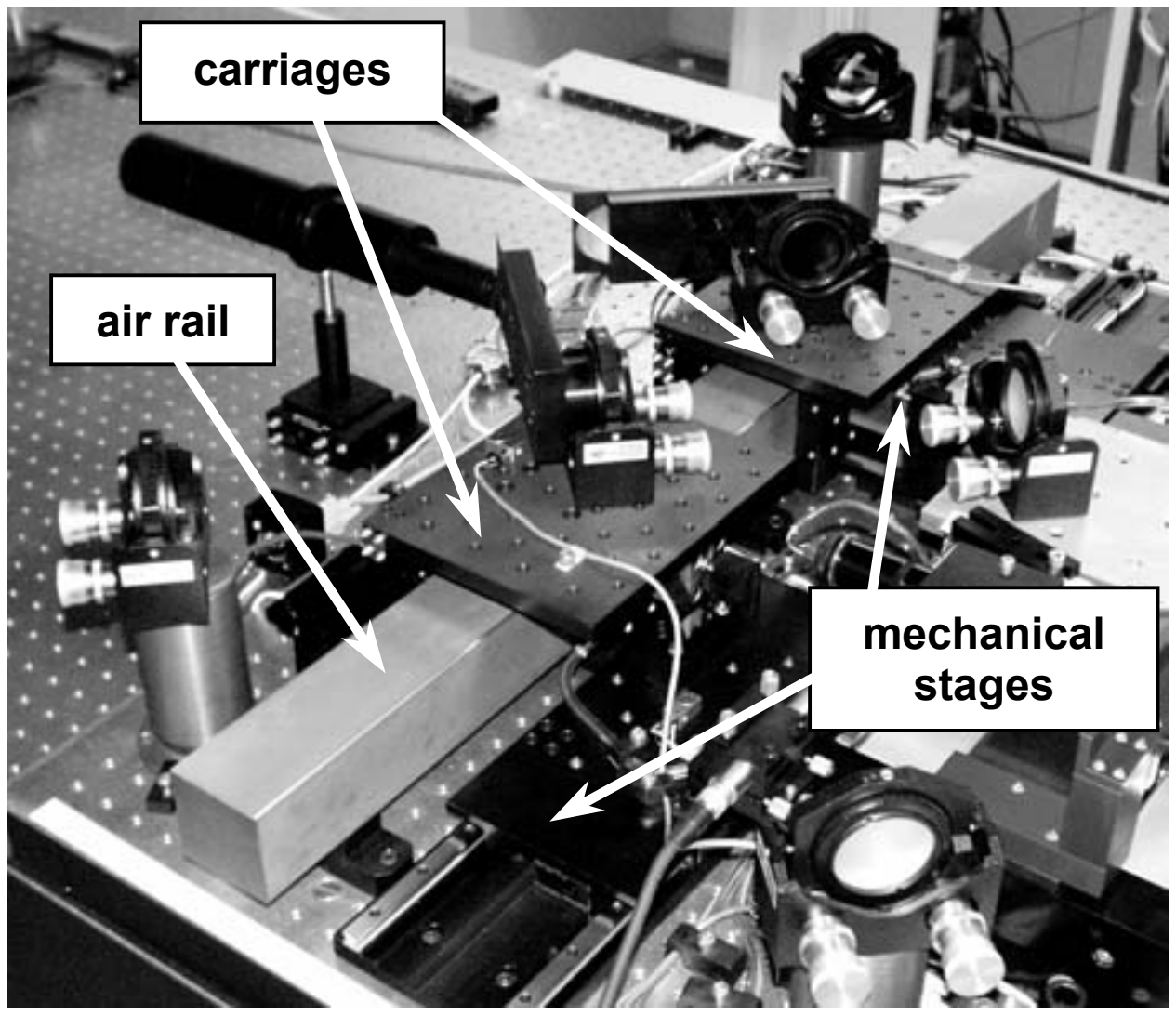

Figure 3 – dual carriage linear air bearing slide driven by mechanical stages with side drive carries baseline mirrors

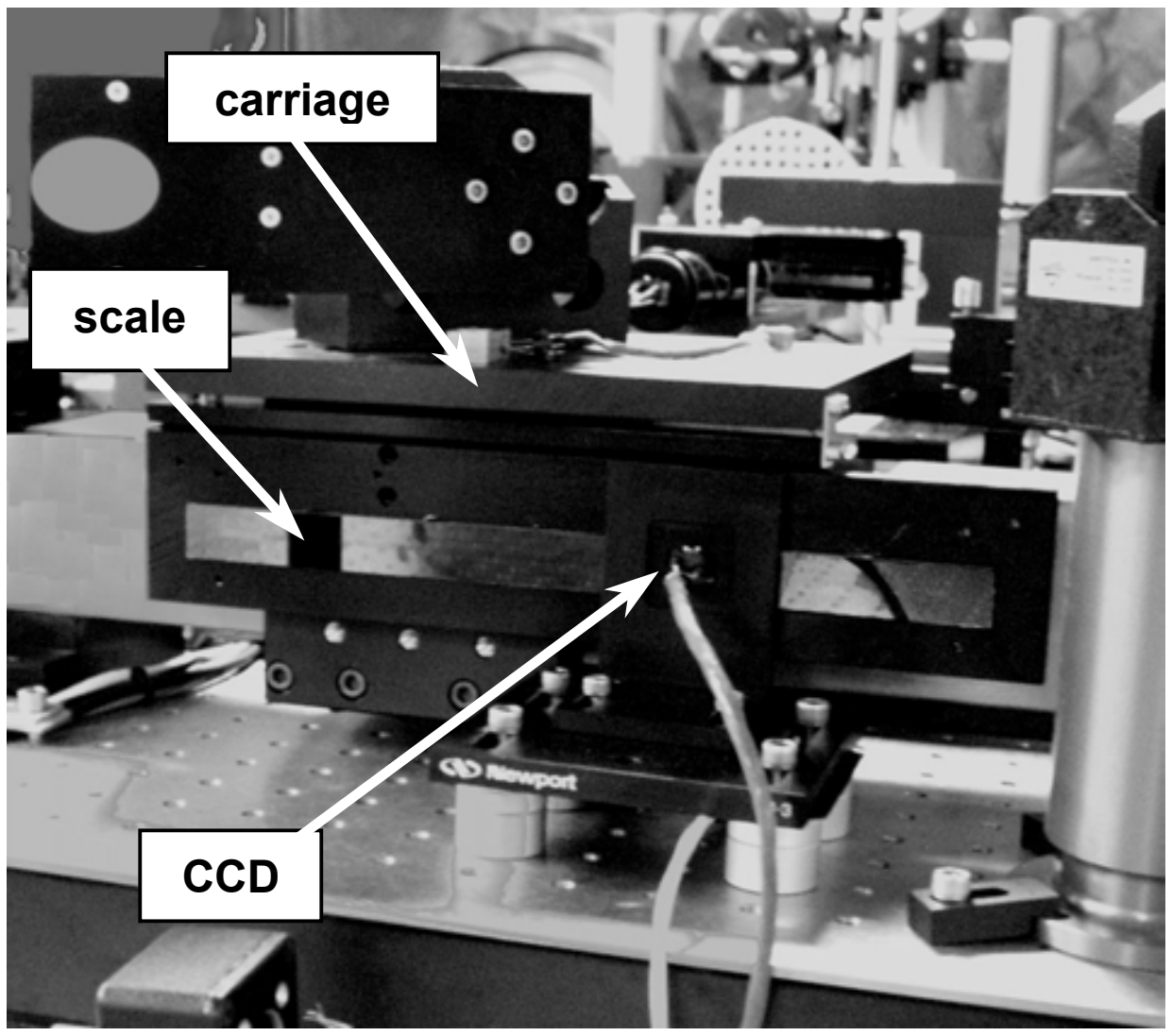

Figure 4 – NASA absolute optical encoders on carriages consist of CCD receiving shadow of scale backlit by LED

Figure 5 shows baseline encoder error versus a DMI for a typical 4 mm portion of travel. Periodic structure at 700 μm is associated with the encoder scale's pitch (major gridlines on the abscissa) and is static. The amplitude of this approximately sinusoidal error is 0.1 μm peak, and lends itself easily to calibration. Meanwhile, because the position reported by the encoder is the result of an image processing algorithm, such errors can be eliminated through modifications to the algorithm once the flaw in the algorithm is identified. This fact notwithstanding, the basic performance of the baseline encoders meets the position knowledge requirement stated above without calibration. The error curve's hairy appearance at high spatial frequency with amplitude of 0.15 μm peak-to-peak is believed to be due to servo jitter of the baseline stage as it is scanned.

As will be seen in a later discussion of errors in the delay line position servo, in addition to the encoders that are built into the mechanical stages only being incremental, those encoders would not be accurate enough to meet the baseline mirror position accuracy requirement in the absence of phase reference sources in the field of view. The use of the NASA absolute encoders make this moot. Operationally, the desired position of each baseline mirror is commanded iteratively until the absolute encoders are satisfied. Then, the DC servo for each axis is commanded to hold whatever current position has been arrived at. The bandwidth requirement for setting the baseline mirror positions is quite low, so this method of closing the loop on baseline mirror positions is acceptable.

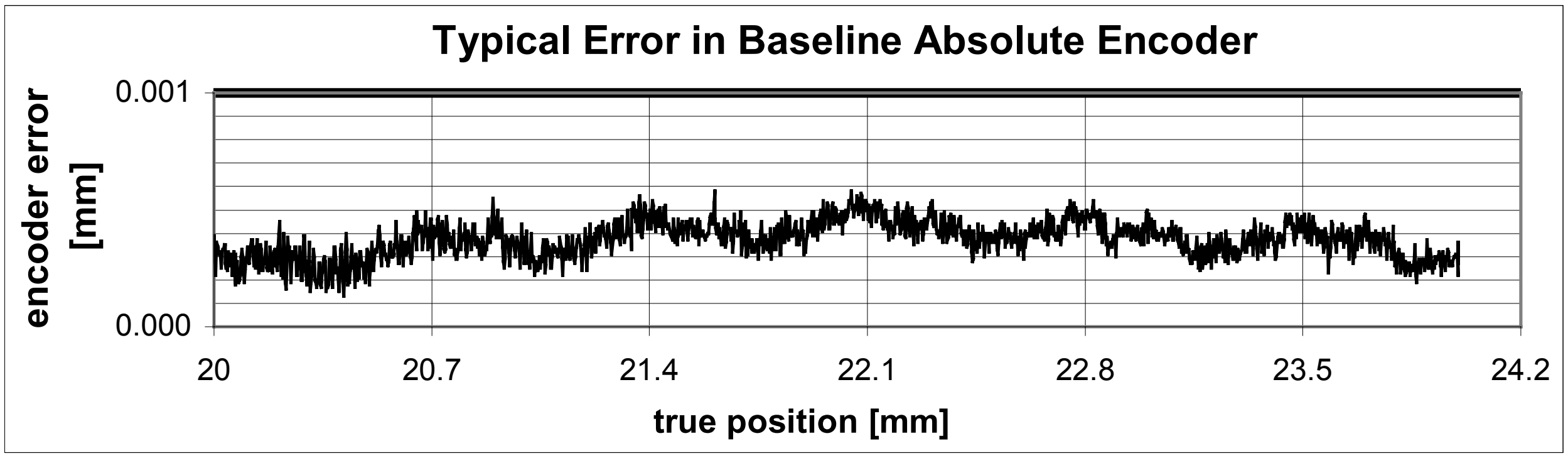

Figure 5 – typical 4 mm section of the uncalibrated error of the absolute encoder on one of the baseline stages.

### 3.2.2 Delay Line

Following the diagonal mirrors on the baseline in each arm are a series of fixed flat mirrors arranged to recombine the two beams at the beamsplitter. In one arm, a pair of flat mirrors arranged as a dihedral move together on a DC servo-controlled air bearing slide which provides optical delay between the two arms. The method of operation of the baseline mirrors and delay line in consort is presented elsewhere and in a companion paper[8]. The metrology requirements for the delay line are the most difficult and critical for the testbed and so warrant more detailed discussion.

The requirement on knowledge of the absolute position of the delay line had at one time been believed to be 0.1 μm, and the delay line stage had been ordered to have that resolution, even though absolute encoders were not available on the stage.[5] Subsequently, it was determined that the requirement, derived from the desired image quality of the testbed (predicated on absolute knowledge of the relationship of ZPD's for all different baseline settings) and the testbed's signal-to-noise goal in the raw, fringe power measurement for a simple sine wave fringe, was actually a much tighter 0.02 μm (one thirtieth of a wavelength). The tightness of this requirement is exacerbated by the fact that the dihedral shifts the phase of light passing through it by twice the stroke of the delay line stage.

Meanwhile, in early tests of interferometric fringe capture with the delay line servo-ing at the 0.1 μm level but not moving, it was found that the delay line jittered at up to that level depending on exactly where the delay line stage was positioned along its travel. This jitter was also clearly evident in visual examinations of the straight fringes in the pupil plane for a point source with a slight tilt introduced between the two arms. Fringe data captured with the delay line moving and with a laser as the source showed wildly erratic departures from the anticipated, smooth, sinusoidal fringe envelope. It was realized that delay line servo jitter and servo velocity errors were at fault.

In addition to requiring higher resolution for position knowledge on that stage and in hopes of mitigating the jitter problem, the stage's built in encoder was upgraded to 0.02 μm resolution through purchase of a five times higher electronic interpolator for its quadrature signals. Initially, from visual inspection of pupil plane fringes, jitter appeared to be markedly improved as a result of the upgrade. At the same time, a plan was developed to use a surplus DMI to generate trigger pulses for fringe image capture at very regular intervals, specifically on logical transitions of one of its quadrature signals, which could be set to occur at a selectable spatial resolution. This would enable straightforward spectral analysis of fringe data through fast Fourier transform. Current indications from trials are that this method will work very well. Still other improvements to the testbed in the form of shielding beams from air turbulence, etc., are believed to be necessary to optimize fringe quality.

It is also desirable to have absolute knowledge of delay line position through absolute encoders like those on the baseline stages, but at a far higher resolution. The main advantage to using absolute encoders on the baseline and delay line stages versus the incremental ones built into the stages is that once ZPD is found, it can easily be found again and again by appealing to the absolute readings of the encoders, even for semi-arbitrary positions of the baseline mirrors.

Figure 6 shows the configuration of the delay line's absolute encoder. Mounting the encoder's components with high stability presents interesting challenges. In this case, the encoder scale, whose fiducial pitch is 60 μm, is mounted rigidly to the base of the delay line stage with stout supporting parts. The CCD array behind a 10X microscope rides on the stage's traveling air carriage. A strain-relieved service loop is provided for the CCD's wiring. An ultra-bright LED, which moves with the microscope, backlights the scale from a distance of about 35 mm to avoid interference with the scale holder. The depth of focus of the microscope is 25 μm. The scale holder has precise adjustments for roll and yaw so that the patterned face of the encoder scale can be adjusted to be exactly parallel to the direction of travel of the air bearing so that the scale will stay in focus over the full 60 mm range of the scale.

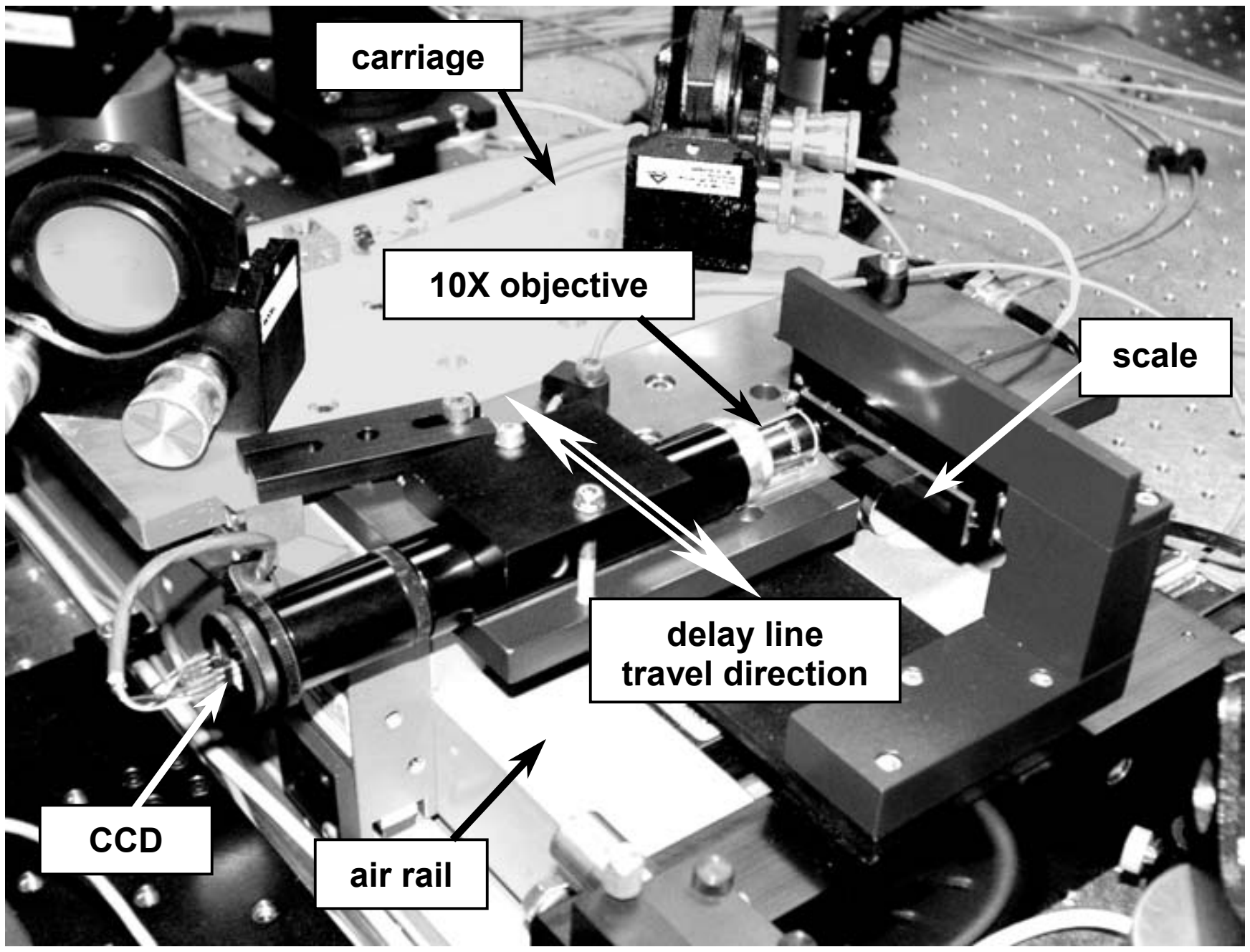

Figure 6 – absolute encoder for delay line air bearing stage has fixed glass scale viewed by traveling 10X CCD microscope

Figure 7 shows the detrended difference between the encoder and a DMI over the length of the encoder scale in 2 mm increments. Linear detrending is necessary because a) due to thermal expansion, the length and pitch of the encoder scale are different at the temperature in the lab versus the temperature at which the scale was made, and b) the DMI is not environmentally compensated, so the DMI's wavelength in the lab air is different from its calibration wavelength. The residual difference is of the order of 0.2 μm peak-to-peak over the full range of the encoder.

At the juncture when both the DMI and the absolute encoder were added to the delay line stage, scans were performed to examine servo velocity performance. The delay line was commanded to scan at constant velocity over a distance of about 200 μm. Encoder data was gathered at some constant frequency of roughly 8 Hz. Each encode triggered a reading from the DMI. Figure 8 is a plot of servo position error as a function of distance along the scan assuming constant velocity. The error assessed by the encoder (top curve) is nearly identical to that seen by the DMI (middle curve) at each point in the scan. The lower curve is the difference between the two. The servo error appears to be nearly sinusoidal over the interval with an amplitude of 0.9 μm peak-to-peak (about 1.5 visible wavelengths) and a period of 20 μm. The servo position error is presumably interpolation error associated with the stage's incremental encoder whose pitch is 20 μm (the same as the major gridlines on the abscissa).

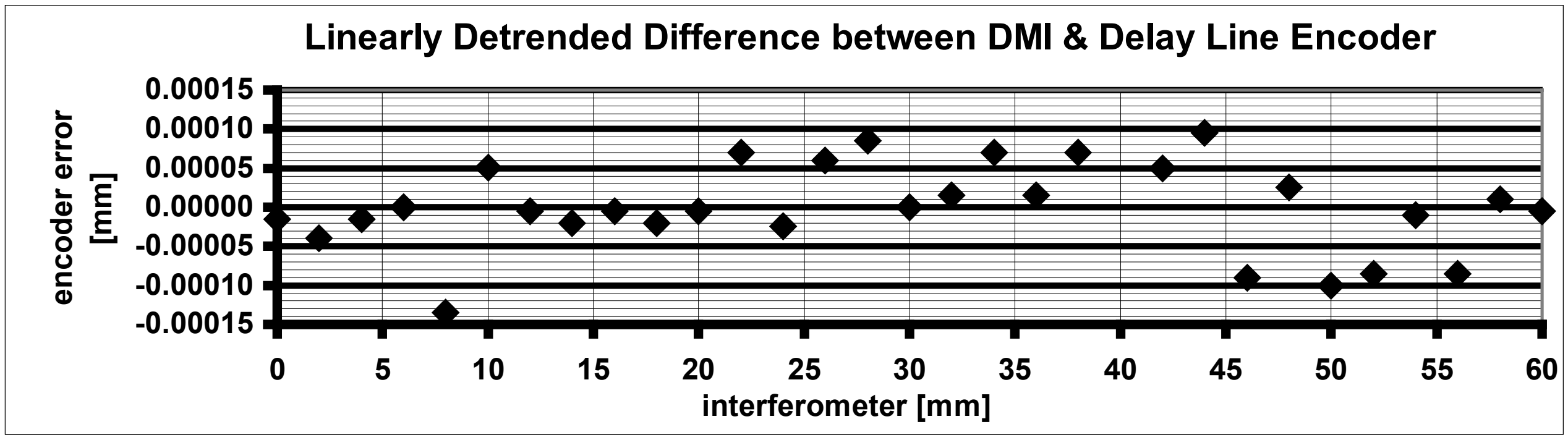

Figure 7 – uncalibrated error of delay line absolute encoder over its 60 mm range

The absolute encoder and the DMI agree with each other for this short scan to a level of just over 0.1 μm peak-to-peak. The fine structure on the difference curve is due to servo jitter, and owes to temporal latency between the encoder and DMI readings and the fact that the jitter bandwidth is much faster than the encoder data acquisition bandwidth. The light gray arrows which are spaced by the incremental encoder's scale pitch point out regions where the fine structure in the difference nearly vanishes. These are believed to be at points within the servo's period where jitter is well behaved. Half way between the arrows, the fine structure grows to maximum amplitude where the servo jitter is not well behaved.

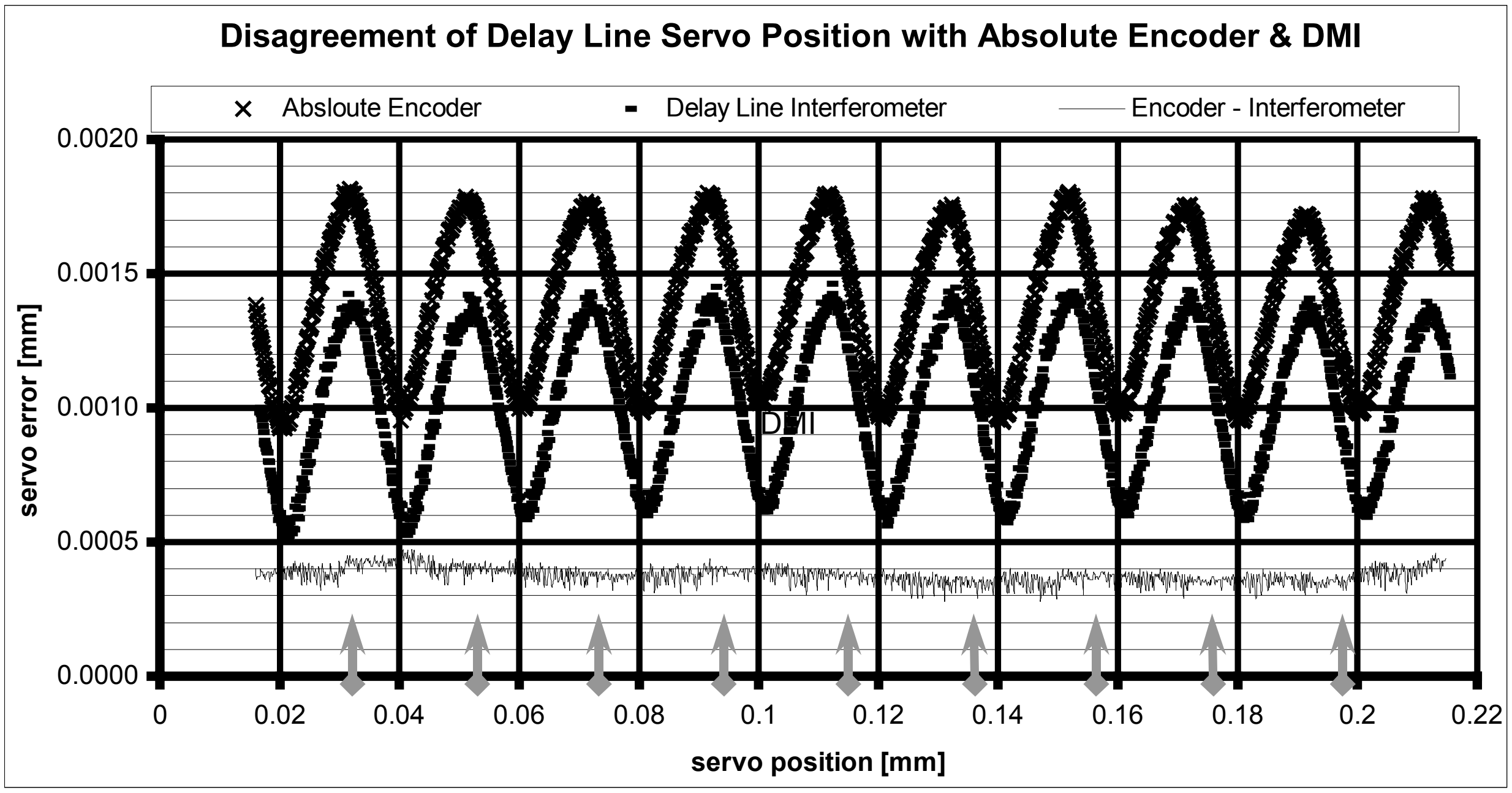

Figure 8 – disagreement between servo position and positions reported by absolute encoder and interferometer

### 3.3 Detector assembly

Like the source assembly, the detector assembly is built on a indexing air bearing spindle whose axis, in this case, is aligned along the optical axis of the re-imaging, achromatic doublet following the beamsplitter. The CCD camera is mounted to the end of a long baffle tube that excludes ambient light from the otherwise exposed CCD array in the camera housing and limits what the CCD can view to what comes through the lens. Because the purpose of the detector's rotation stage is to de-rotate the source field for certain testbed observations, the requirement on knowledge of absolute angle of detector rotation is the same as that for source rotation – 0.3 arcseconds. In fact, the rotation stage for the detector is a twin of the source rotation stage. The camera is mounted on the rotation stage so that the axis of rotation passes through a pixel near the center of the CCD array. The absolute encoder configuration for the detector assembly is shown in Figure 2c.

## 4. TESTBED OPTICAL ALIGNMENT

In this section, the mutual, opto-mechanical co-alignment of assemblies and components within assemblies are described. Most of these alignments rely on straightforward optical alignment procedures. Having an accurate CAD model of the testbed components is of great benefit in these alignments because there is some assurance that everything in the crowded testbed will fit in the allowed real estate and placement of components goes quickly when things are designed to be placed over specific tapped holes in the surface of the optical table. The chief goal is to get the beams of the two arms of the receiver to interfere at the beamsplitter at right angles without shear (for maximum fringe contrast).

### 4.1 Collimator and source

First, for convenience, the axis of the collimating mirror is roughly aligned to be parallel to the surface of the optical table and parallel to the rows of tapped holes in table. To start, a crosshair reticle is set up over the table at the nominal source location which is to be at the collimator's focal point. Next, crosshairs are strung across the collimator's aperture and the collimator position is coarsely adjusted to be at the same height off the table as the reticle and over the same row of holes. The crosshair over the mirror now serves as a target for a second point on the nominal opto-mechanical axis.

An alignment telescope is adjusted so that it can alternately focus on the reticle and the crosshair on the collimator, so that its axis defines the intended axis of the collimator. Once that axis is established, a transfer flat is arranged so that it can now be seen in autocollimation by refocusing the alignment telescope to infinity focus. (By transfer flat, we mean a flat mirror which is wide enough to span a lateral gap between to lines of sight that are to be established parallel to one another. Such a flat allows an alignment telescope to be set up in autocollimation then moved laterally and set up again in autocollimation to re-establish its original pointing direction.) The surface of the transfer flat is now normal to the nominal opto-mechanical axis. The alignment telescope is moved laterally so that its projected aperture is shared by the transfer flat and an alignment reference cube epoxied to the edge of the collimator's aperture. One face of the cube is nearly perpendicular to the collimator's optical axis with small but known offsets in tip and rotation. The alignment telescope is re-adjusted to be again in autocollimation with the transfer flat. Finally, collimator tip and rotation are adjusted so that the cube is seen at the appropriate angular offsets with respect to the transfer flat. The collimator axis is now quite parallel to the table and the reticle is nearly at the focal point of the collimator. The transfer flat is now removed so that the alignment telescope, still focused at infinity, views the source reticle through the collimator. The position of the reticle, on an X-Y-Z micrometer stage, is now adjusted to appear on the crosshair of the alignment telescope and focused as well as possible visually. The reticle is now replaced by a source pinhole which serves as a spatial filter for a laser. Focus is set by optimizing shearing interferometer fringes from the laser in collimated space.

### 4.2 Source spindle axis to collimator axis

Aligning the source spindle to the axis of the collimator is actually somewhat entertaining. Once the whole source assembly has been positioned so that the glass target array surface is in the focal plane of the collimator and the alignment telescope is trained on the focal point, the axis of the spindle can be found easily and then made coincident with the collimator axis. First, one rotates the spindle and identifies the target feature which appears to travel on a circle with the smallest radius. One then adjusts the X-Y target adjusting micrometers so that that feature appears not to rotate at all. That feature is now on the rotation axis of the spindle. Now, the entire source assembly is positioned as a unit until that same feature is seen on the crosshairs of the alignment telescope. The axis of the spindle is now coincident with the collimator axis.

### 4.3 Receiver mirrors

The procedure for adjusting each receiver mirror uses similar techniques to those used in aligning the source reticle to the collimator axis. The process involves alternately referring to an autocollimator's view of a) the source seen through the collimating mirror, b) the source seen through the receiver mirror which is being aligned, c) the optical reference cube on the side of the collimator, d) or combinations of these seen either directly or through the use of a transfer flat. One additional alignment tool, a "true square" is used when the beam direction of a receiver mirror is to be folded to travel perpendicular to the collimator axis. A true square is a polished metal gauge block whose faces are optically flat, extremely perpendicular (arcsecond level), and calibrated in relative angle to a very high accuracy (<0.1 arcsecond). Although true squares come in various sizes, the one we use has faces which are 50 mm wide and 25 mm high.

The mirrors in each arm are aligned in sequence starting with diagonal mirrors on the baseline. First, the autocollimator is set up to view the source seen through the collimating mirror on its crosshairs, looking just past the back of each diagonal mirror's mount. Next, the true square is positioned on the baseline carriage so that its side face will share the autocollimator aperture of about 125 mm diameter with the source beam folded by the diagonal mirror. The true square is adjusted to be in autocollimation (viewed on the crosshair of the autocollimator). The autocollimator is now moved and adjusted to view the side face of the true square in autocollimation. The tip and rotation of the diagonal mirror is now adjusted so that the source pinhole is seen on the crosshairs. The source beam has now been folded to be perpendicular to the optical axis of the collimator. This process is illustrated schematically in Figure 9a.

The next fold mirror in that arm is now adjusted to make the beam parallel to the axis of the collimator. But first, the position of the mirror itself must be assured to be mechanically centered on the 25 mm diameter beam sampled by the baseline mirror, preventing vignetting of the beam as it travels through the arm. This is done by eye, looking back into the mirror, simply making sure that the beam appears to be centered, that is, that the source appears to be extinguished sharply at the same radius in the aperture as the eye is moved transversely to the folded beam. Next, the autocollimator is moved to once again simultaneously view the source directly and through the fold mirror (if possible) on the autocollimator's crosshairs. The tip and rotation of the fold flat is now adjusted so that the source pinhole is seen on the crosshairs. It is sometimes necessary to use the transfer flat to achieve this desired coincidence of pointing (Figure 9b).

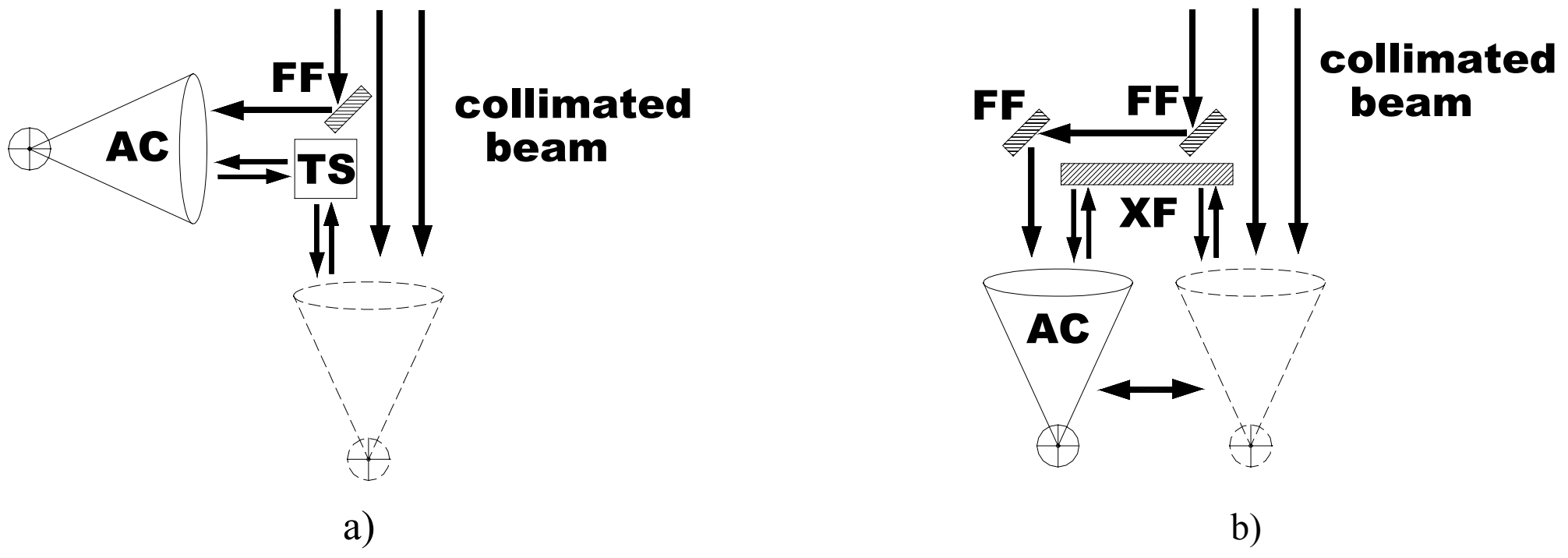

Figure 9 – a) alternately viewing the source directly with an autocollimator (AC) and through a diagonal fold flat (FF) aided by a true square (TS); b) using a transfer flat (XF) to co-align a receiver beam to the collimated beam

### 4.4 Beamsplitter

The sequential process of aligning one arm and then repeated for the other arm. Eventually, the two beams are found to overlap at the location where the beamsplitter will be placed. If the alignment has gone to plan to this point, there will be a location where the circular cross-section of the two beams will overlap with no shear. This is where an ideal, infinitely thin beamsplitter would be placed. Its body orientation is adjusted so that the autocollimator's view of the source emanating from one arm, say, the one transmitted by the beamsplitter, is coincident with the view of the source from the other arm reflected by the beamsplitter. This is, in fact, what the testbed's detector sees under normal conditions – a single image of the source magnified by the ratio of focal lengths of the re-imaging lens and collimator, respectively (roughly 0.31). Once achieved, the view through the other beamsplitter output path looks just the same.

With our slab type beamsplitter, there is an equivalent location inside the volume of the optic where each beam is offset equilaterally by refraction at the input faces. Nonetheless, the adjustment process is the same as outlined for the ideal beamsplitter. The beamsplitter is mounted on top of a ultra-high precision prism table which has pitch, roll, and yaw adjustments for achieving the required body orientation of the optic. The prism table is in turn mounted on an X-Y micrometer stage so that the shear between the two input beams can be eliminated at the beam combining surface.

With a pinhole as a source and coherent source light, one can observe classic fringes from the interference of two plane waves on a diffuser (such as ground glass) placed just beyond either beamsplitter output face. When the collimated beams from the two arms exit the beamsplitter in the same direction, a null fringe is observed. By conservation of energy, when maximum fringe brightness is seen through one exit face, minimum brightness is seen through the other.

If one tilts the beamsplitter slightly, straight and parallel fringes are observed, because the beams transmitted and reflected by the beamsplitter no longer travel in the same direction. At the same time, two separate spots appear in the view through the autocollimator or in the image recorded by the detector (Figure 10). In practice, it is the fine alignment of the beamsplitter which assures that the criterion outlined in the second paragraph of section 3.2.1 for re-imaging a point s

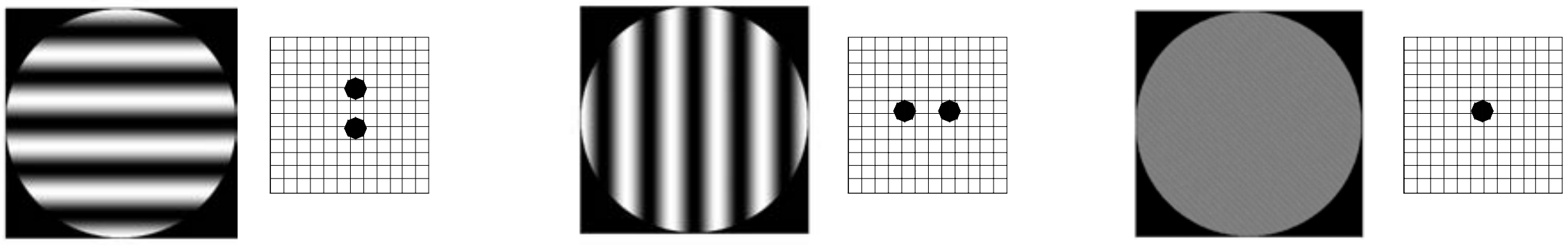

a) beamsplitter tip b) beamsplitter rotation c) "nulled" beamsplitter

Figure 10 – schematic views of fringes observed on diffuser through beamsplitter exit faces and corresponding images on detector

### 4.4 Detector

The imaging lens is diffraction-limited even for off-axis light well outside of the field of view of interest to the testbed, so aligning the lens axis to the direction of the on-axis beam exiting the beamsplitter is fairly trivial. Of the six available degrees of freedom for detector alignment, the two Cartesian degrees of freedom for decenter transform to spindle rotation during observation and radial position of some arbitrary detector pixel relative to the spindle axis. The long baffle tube screwed into the detector housing is mounted to the spindle with a precision collet. The transverse position of the collet on the spindle's worktable determines the detector pixel through which the spindle axis passes. The latter adjustment is made by looking at the detector surface from collimated space with the autocollimator and assuring that the detector appears to rotate about a point near its center. Adjustments to the position of the detector assembly as a whole now allow the image of an on-axis point source to be made to coincide with that particular pixel. The only remaining adjustment is focus. Fine focus is achieved by slipping the baffle tube axially in the collet until the image of the source is as sharp as possible. Mechanical tolerances of the camera housing, tubing, collet, and spindle worktable are sufficient to assure that the normal to the detector surface is adequately parallel to the spindle axis.

## 5. ENVIRONMENTAL MONITORING

Those experienced in laboratory interferometry are familiar with the effects of turbulent air on interference fringes. In WIIT, turbulence creates gradients in the refractive index of air at many spatial scales, causing undesirable changes in the optical paths for the two arms of the receiver, producing undesirable fringe modulation. We have taken precautions to minimize the effects of several sources of turbulence including heat from active devices near the testbed, cycling of laboratory air conditioning, fans blowing in nearby electronic equipment, and exhaust air from the testbed's air bearings.

Fringe quality improved markedly after we enclosed the entire testbed in a tent, preventing fans from blowing air across the testbed and reducing the effects of the lab's ventilation systems turning on and off. For the fluorescent work light above the testbed, only its diffuser is inside the tent. The light's heat generating parts are above the ceiling of the tent. The testbed's incandescent white light source, which gets quite hot, sits on top of the work light. Its light is guided to the source plane through a fiber optic bundle. Air heated by these light sources cannot mix directly with the air inside the tent which would create turbulence. The DMI's laser source must be mounted on the optical table, however, it is walled off with sheet insulation in a corner of the table as far from the receiver as possible. The laser beam passes through a small hole in the insulation. Air heated by the laser is exhausted outside the tent.

Occasionally, the same DMI which triggers image capture at extremely regular increments of delay line motion is rearranged to monitor changes in optical path length due to moving air and temperature gradients for other important paths in the receiver. Results of such characterizations are reported in a companion paper.[8] A network of sixteen LM135 temperature sensors distributed across the testbed concurrently track temperatures of various receiver components (active and passive) and locations in the air above the optical table. The LM135 was selected for its simplicity of operation, ease of calibration, and rated temperature resolution and long term stability. Its output is a voltage which when multiplied by 100 reads out reads out directly in Centigrade. Temperature readings of 0.02 C resolution are logged by reading the output voltages from the sensors through a sixteen channel multiplexer for a 16 bit analog-to-digital converter. Sensor locations are listed in Figure 11a. Figure 11b is a photo of a sensor in its copper block mount. Figure 11c is a plot of testbed temperatures for a 24 hour period in which people were working on the testbed. Sensors in air respond more quickly to temperatures changes than do those mounted to more massive testbed components. The plot

**0 Source, low**
**1 Source, high**
**2 Camera, low**
**3 Camera, high**
**4 Collimator, low**
**5 Collimator, high**
**6 Receiver mirror**
**7 Baseline stage, L**
**8 Receiver mirror, R**
**9 Baseline stage, R**
**10 Delay line, dihedral**
**11 Delay line, stage**
**12 Beamsplitter**
**13 Baseplate, aft**
**14 Baseplate, center**
**15 Baseplate, forward**

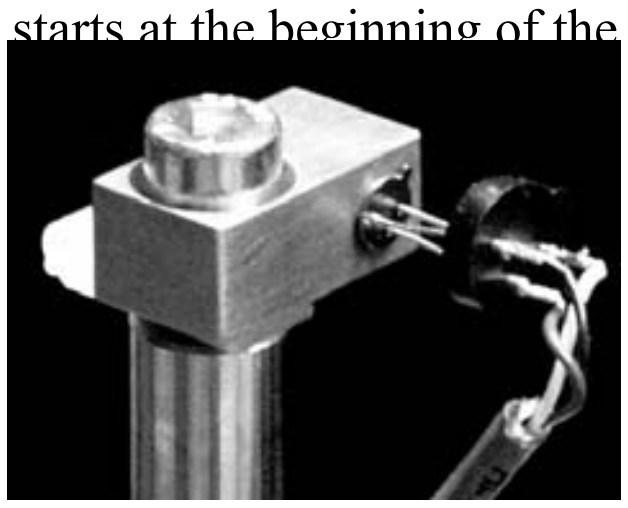

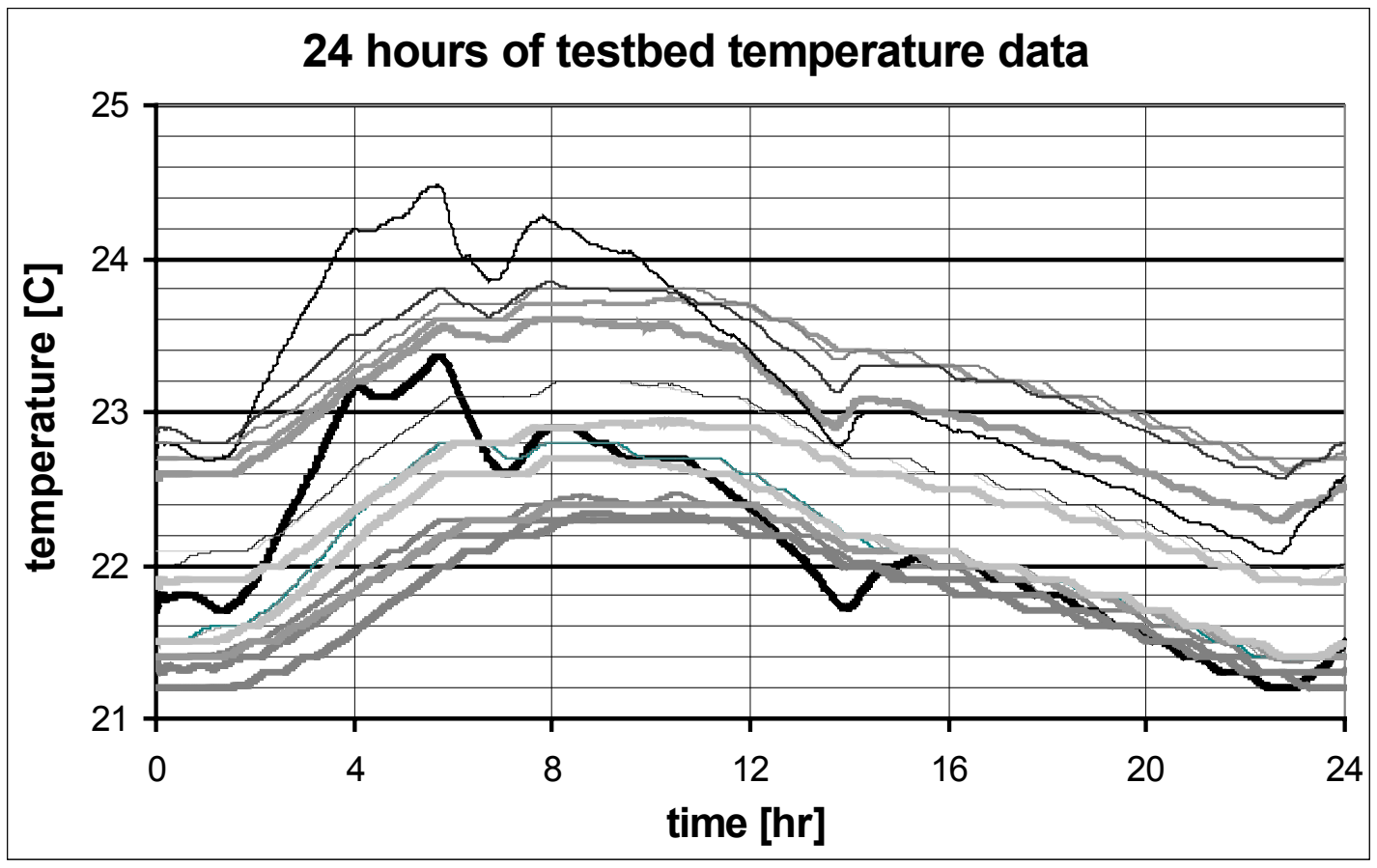

a) b) c)

Figure 11 – a) temperature sensor locations; b) temperature sensor in copper block mount; and c) 24 hours of logged temperature data

## 6. OPTICAL POWER MONITORING

The scene being observed by WIIT must appear to be constant in brightness over the course of interferometric observations of that scene. Unaccounted for changes in source brightness result in errors in synthesizing the scene from image data. A silicon photodiode connected to one of two channels of an IEEE-488 controlled optical power meter intercepts a portion of the collimated beam just below the baseline mirrors at the center of baseline travel and monitors brightness of the source during observations (Figure 12). Brightness fluctuations with bandwidths of less than 2 Hz are tracked. Image data are n alysis.

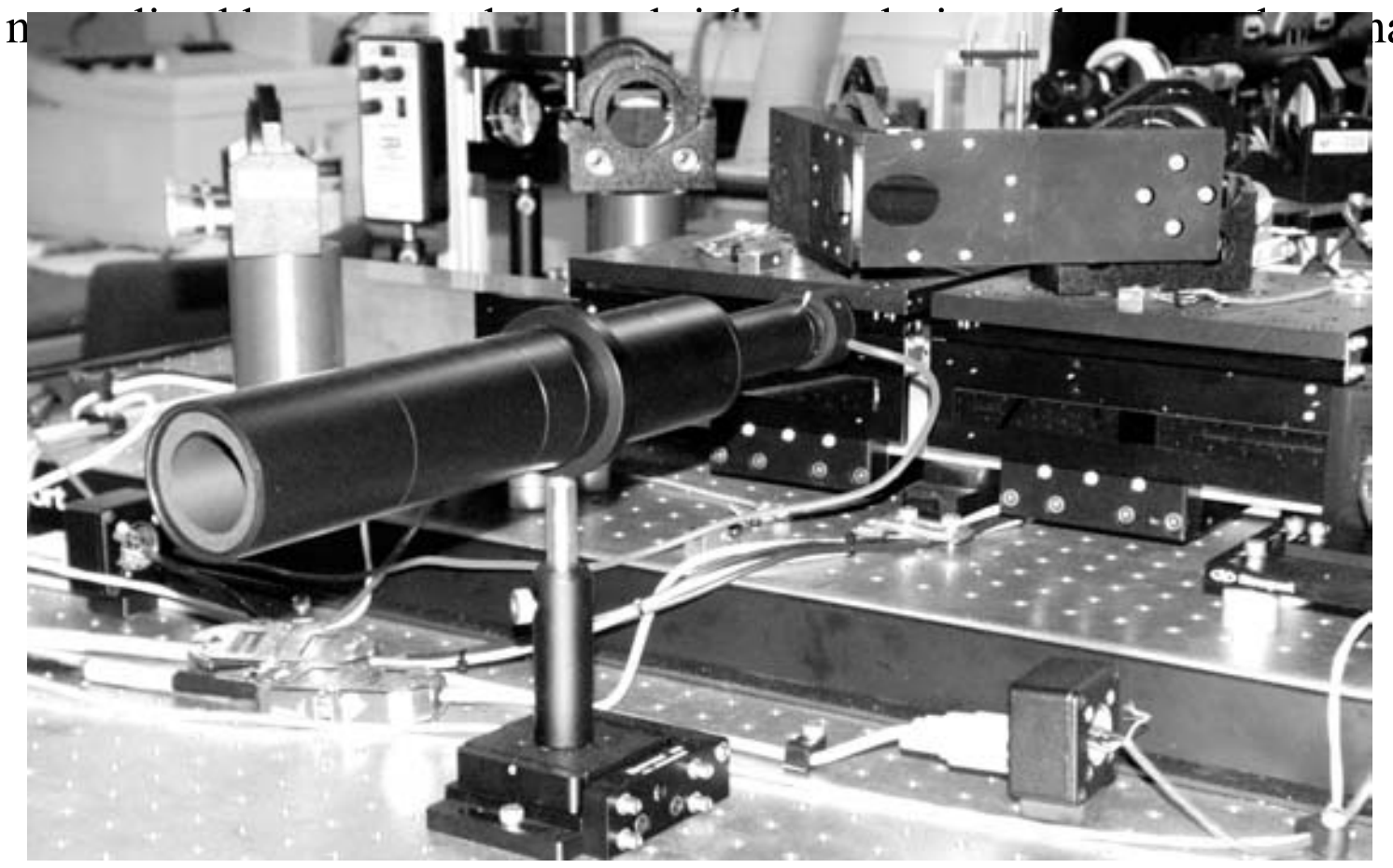

Figure 12 – silicon photodiode radiometer below receiver baseline mirrors is baffled to monitor brightness of source light only

It is also important for all portions of the u-v plane to appear to be equally bright as they would from space. Differences in optical power coupled through each receiver arm affect fringe contrast at the detector. A second photodiode located at the unused beamsplitter exit face is connected to the second channel of the power meter, and the ratio of brightness readings made with each arm alternately blocked for all combinations of baseline mirror positions used during an observation are used to model effects on fringe contrast in subsequent analysis of image data.

## 7. FUTURE PLANS

Most of the metrology tasks described here require only low bandwidth. Source and detector rotations and baseline settings are made and then image data is collected as the delay line is scanned for many minutes at a time. Temperatures and source brightness change slowly. The DMI is autonomous. The highest bandwidth metrology task is reading the absolute encoder on the delay line stage to occasionally corroborate the implicit position of the delay line due to correct triggering of camera frames by the DMI. In time, the delay line encoder will be replaced with a high speed version which will be able to report absolute position at the same rate as the DMI[9]. All other aspects of control and data acquisition for WIIT such as image capture and data logging, and delay line scan control are very computation intensive. Therefore all of the low bandwidth tasks discussed are relegated to a separate computer which receives commands to act and report through file interchange via a shared drive on a local area network. The overall electronic architecture for WIIT and the inter-relationship of controllers are discussed in a companion paper[8].

Our near term plans involve improving the immunity of the testbed to environmental effects. Both the beams in each arm of the receiver and the beam for the DMI will be enclosed in an attempt to extinguish turbulence effects. Meanwhile, it has become clear that a more sophisticated absolute metrology system will be needed in case the strategy to appeal to in-field sources for ZPD phase reference for fringe data taken at different baseline settings turns out to be impractical. This will require well-calibrated, absolute baseline metrology significantly below the current 0.05 μm capability. It is expected that some combination of DMI's and absolute encoders of the style which has been installed on the delay line will have to be applied to the baseline stages as well. Our main mid-range plan for improvements to metrology and control of image capture include the implementation of a new two channel DMI system.

In contrast to the current DMI, the readout of each channel of the new DMI is numeric. Instead of just measuring the stroke of the delay line alone, we plan to use both channels to measure changes in the total lengths of each receiver arm not only for environmental diagnostic value, but also for the purpose of triggering image captures based on real time incrementation of the difference of the lengths of the two arms. The DMI beams will double pass the receiver arms using unilluminated patches of the receiver mirrors starting at the beamsplitter and terminating at flat retroreflectors on the baseline mirrors. This should optimize knowledge of effective optical delay with the hope of eliminating fringe modulation due to unintended changes in optical path length of each arm. This may also provide suitable ZPD phase reference for all baselines. Longer range plans involve possible real time optical monitoring and correction of the pointing direction of the baseline mirrors as they are moved along the air bearing rail to optimize image coincidence on the CCD from both arms of the receiver.